\documentclass[sigconf,anonymous=false]{acmart}
\AtBeginDocument{%
  }

\copyrightyear{2026}
\acmYear{2026}
\setcopyright{cc}
\setcctype{by}
\acmConference[WWW Companion '26]{Companion Proceedings of the ACM Web Conference 2026}{April 13--17, 2026}{Dubai, United Arab Emirates}
\acmBooktitle{Companion Proceedings of the ACM Web Conference 2026 (WWW Companion '26), April 13--17, 2026, Dubai, United Arab Emirates}
\acmDOI{10.1145/3774905.3794700}
\acmISBN{979-8-4007-2308-7/2026/04}
\usepackage [
lambda,
advantage,
operators,
sets,
adversary,
landau,
probability,
notions,
logic,
ff,
mm,
primitives,
events,
complexity,
asymptotics,
keys
] {cryptocode}

\usepackage{verbatim}
\usepackage{pgfplots}
\usepgfplotslibrary{statistics}
\usetikzlibrary{math}

\pgfplotsset{width=\columnwidth,compat=1.9} 

\createprocedureblock{interactGame}{center,boxed}{}{}{}

\NewDocumentEnvironment{gameInteract}{m+b}
 {%
  \interactGame[linenumbering]{#1}{#2}
  \par\addvspace{\baselineskip}
 }{}

\usepackage{amsmath}
\usepackage{amsfonts}
\usepackage{algorithm}
\usepackage[noend]{algpseudocode}
\usepackage{xcolor}

\newcommand{\Abs}[1]{\left| #1 \right|}

\newcommand{\Norm}[1]{\ensuremath{\left\lVert #1 \right\rVert}\xspace}

\usepackage{xspace}

\newcommand{\R}{\mathbb{R}}

\newcommand{\E}[1]{\mathbb{E}\left[ #1 \right]}

\newcommand{\B}{\ensuremath{\mathbf{B}}\xspace}

\newcommand{\MI}[1]{\ensuremath{\mathrm{MI}\left( #1 \right)}\xspace}

\newcommand{\M}[1]{\ensuremath{\mathcal{M}\left( #1 \right)}\xspace}

\newcommand{\Distrib}[1]{\ensuremath{\mathsf{D}}\xspace}

\newcommand{\Sig}{\ensuremath{\Sigma}\xspace}

\renewcommand{\E}{\ensuremath{\mathbb{E}}\xspace}

\renewcommand{\epsilon}{\varepsilon}

\newcommand{\domain}{\ensuremath{\mathcal{X}}\xspace}

\newcommand{\Prv}{$\mathcal{P}$\xspace}
\newcommand{\Vrf}{$\mathcal{V}$\xspace}
\newcommand{\Cir}{$\mathcal{C}$\xspace}

\title[PAC to the Future]{PAC to the Future: Zero-Knowledge Proofs of PAC Private Systems}

\begin{document}

\title[PAC to the Future]{PAC to the Future: Zero-Knowledge Proofs of PAC Private Systems}


\author{Guilhem Repetto}
\affiliation{
  \institution{École Normale supérieure of Rennes}
  \city{Bruz}
  \country{France}}
\email{guilhem.repetto@ens-rennes.fr}

\author{Nojan Sheybani}
\affiliation{%
  \institution{University of California, San Diego}
  \city{La Jolla}
  \country{USA}}
\email{nsheybani@ucsd.edu}

\author{Gabrielle De Micheli} 
\affiliation{%
  \institution{University of California, San Diego}
  \city{La Jolla}
  \country{USA}}
\email{gdemicheli@ucsd.edu}

\author{Farinaz Koushanfar}
\affiliation{%
  \institution{University of California, San Diego}
  \city{La Jolla}
  \country{USA}}
\email{farinaz@ucsd.edu}





\renewcommand{\shortauthors}{Repetto et al.}

\begin{abstract}
Privacy concerns in machine learning systems have grown significantly with the increasing reliance on sensitive user data for training large-scale models. This paper introduces a novel framework combining Probably Approximately Correct (PAC) Privacy with zero-knowledge proofs (ZKPs) to provide verifiable privacy guarantees in trustless computing environments. Our approach addresses the limitations of traditional privacy-preserving techniques by enabling users to verify both the correctness of computations and the proper application of privacy-preserving noise, particularly in cloud-based systems. We leverage non-interactive ZKP schemes to generate proofs that attest to the correct implementation of PAC privacy mechanisms while maintaining the confidentiality of proprietary systems.
Our results demonstrate the feasibility of achieving verifiable PAC privacy in outsourced computation, offering a practical solution for maintaining trust in privacy-preserving machine learning and database systems while ensuring computational integrity.
\end{abstract}


\maketitle

\section{Introduction}





The amount of sensitive user data that is required to assist in the technological advances of large-scale machine learning paradigms, such as large-language models \cite{naveed2023comprehensive}, has grown in recent times. This has led to the general public developing concerns about the safety of their data and has caused a general lack of trust when using proprietary systems \cite{jenks2025communicating}. Probably Approximately Correct (PAC) Privacy \cite{xiao2023pac} offers a framework for providing provable privacy guarantees for black-box algorithms. It offers a compelling alternative to traditional noise-based privacy techniques like differential privacy. 

Unlike differential privacy, which can lead to significant deterioration of utility when aiming for strong privacy guarantees \cite{blanco2022critical}, PAC privacy provides rigorous privacy guarantees for black-box algorithms while maintaining a better balance between privacy and utility. Furthermore, PAC privacy's flexibility in determining noise perturbation for any given privacy level makes it easier to adapt to specific application needs, offering a more generalizable and practical framework for maintaining data privacy. While this technique is valuable, ensuring that the appropriate privacy-preserving noise is computed and applied correctly remains a challenge, particularly when the underlying data and execution details must remain confidential.

While innovations in privacy-preserving computing have enabled computations on encrypted or masked data, this approach does not provide users a way to ensure that operations are being computed correctly, or even that the input or obtained data are correct. Zero-knowledge proofs (ZKPs) have emerged as a prime candidate for ensuring privacy \textit{and} integrity in large-scale systems. ZKPs are designed to provide the same privacy guarantees as the prominent privacy-preserving techniques, such as fully homomorphic encryption (FHE) and multi-party computation (MPC), while also providing \textit{verifiability}. 

ZKPs are an excellent solution for enabling private computation in trustless environments, in which the user requests proof that computation is done correctly and securely. This environment is especially prevalent when computation is outsourced to a cloud provider (e.g. ChatGPT). In their current state, besides standard encryption techniques, users of cloud-based systems must trust that computation is sound and secure. Our proposed system aims to address this by using ZKPs in combination with PAC privacy to ensure verifiable privacy in trustless systems.
Our approach combines the rigorous privacy guarantees of PAC Privacy with the verifiability of zero-knowledge proofs, offering a new paradigm for trustworthy privacy-preserving computing. By providing proofs of correct noise application, we enable parties to verify that proper privacy measures have been applied without compromising the confidentiality of a service provider's underlying data.

This paper aims to show the feasibility of achieving verifiable proofs of PAC privacy for outsourced computation. Our system ensures users that their computation is safely secured by PAC privacy, while also allowing them to verify that any results returned by the cloud provider have been correctly computed. Most importantly, our proposed system provides easily verifiable proofs to attest to the correct computation of PAC private noise generation, ensuring to users that the privacy guarantees of PAC privacy are being correctly upheld.

In this paper, we extend the notion of PAC Privacy by introducing a novel approach to verifying the correct computation and application of privacy-preserving noise using zero-knowledge proofs (ZKPs). Specifically, we leverage state-of-the-art zk-STARKs ZKP schemes that allow cloud-based systems to provide proof of correct computation alongside proof that the promised privacy guarantees are being maintained and applied correctly, without revealing any proprietary information. We highlight that this work is built using non-interactive proofs. This is done to allow the generation of publicly verifiable proofs attesting to the correct application of PAC privacy and sound computation.

In short, our contributions are as follows:
\begin{itemize}
    \item We present a novel end-to-end framework for verifiable PAC Privacy with post-quantum secure zero-knowledge proofs in non-interactive settings.
    \item Extensive evaluation of our proposed system on machine learning and database cloud-based operations demonstrates near-plaintext utility while incurring minimal overhead for proof generation.
\end{itemize}

\section{Preliminaries}

\subsection{Zero-Knowledge Proofs}

Zero-Knowledge Proofs (ZKPs) are a cryptographic primitive that allow a prover \Prv to prove to a verifier \Vrf that they know a secret value $w$, often called \textit{witness}, without revealing anything about $w$. Formally, ZKPs allow \Prv to prove to \Vrf that they know a secret input $w$ to a computation \Cir such that $\mathcal{C}(x; w)=y$, where $x$ and $y$ are public inputs and outputs, respectively. ZKPs have been primarily used to allow users to prove knowledge of private data \cite{hasan2019overview} and to prove the \textit{correct} computation of a function with public and/or private data \cite{xing2023zero}, known as verifiable computation. There exist several constructions of ZKPs, two of which we will discuss in further detail, that target different attributes, such as post-quantum security or proof succinctness, at the cost of runtime, trusted setup assumptions, or communication. Despite these different constructions, all ZKPs have three core attributes \cite{goldreich1994definitions}:
\begin{enumerate}
    \item \textbf{Soundness}: \Vrf will find out, with a very high probability, if a \Prv is dishonest if the statement is false.
    \item \textbf{Completeness}: An honest \Prv can convince \Vrf if the statement is true.
    \item \textbf{Zero-Knowledge}: If the statement is true, \Vrf will learn nothing about the \Prv's private inputs.
\end{enumerate}

One of the most mature constructions of ZKPs is Zero-Knowledge Succinct Non-Interactive Arguments of Knowledge (zk-SNARKs) \cite{chen2022review}, which are publicly-verifiable succinct proofs, defined as small proofs (around 128 bytes) that can be verified quickly by any \Vrf. While these have grown to prominence in the blockchain, due to their succinctness, oftentimes zk-SNARKs rely on a trusted setup process for every new computation \Cir which can be computationally heavy and rely on a third party. Alongside this, proof generation in zk-SNARKs is computationally heavy due to the effort required to achieve the succinctness property and the underlying cryptography that is used: elliptic curve cryptography, which is not post-quantum secure. For these reasons, we do not consider zk-SNARKs in our proposed approach, and instead focus on the following state-of-the-art ZKP construction:

\textbf{Zero-Knowledge Scalable, Transparent Arguments of Knowledge (zk-STARKs)} remove the dependence on a trusted setup by using publicly verifiable randomness for generating the parameters for proof generation and verification \cite{sheybani2025zero}. This is referred to as \textit{transparency} in ZK literature. These protocols only cryptographically rely on collision-resistant hash functions, which serve as a lightweight approach toward achieving post-quantum safety. By default, zk-STARKs were formulated as interactive systems, but these protocols can be made non-interactive by applying the Fiat-Shamir transformation \cite{goldwasser2003security}. Technically, a non-interactive zk-STARK can be classed as a zk-SNARK, while a transparent zk-SNARK can be classified as a zk-STARK. Those zk-STARKs feature as the underlying schemes in prominent ZK virtual machines, such as RISC-Zero \cite{Risc0}, which is utilized in our presented work, and SP1 \cite{succinctlabs_sp1}, due to their efficient computational overhead, lack of trusted setup, and generation of publicly-verifiable proofs due to their non-interactive nature.

\subsection{PAC Privacy}



PAC privacy was introduced in~\cite{xiao2023pac}. The main goal of PAC privacy is to avoid public data from leaking sensitive information. The proposed solution publishes a degraded version of this data by adding random noise to it. PAC privacy describes a very general procedure to analyze the data and generate an assorted noise, such that the processed data is still relevant while the private sensitive information is protected.
We recall the important concepts from~\cite{xiao2023pac}:
\begin{itemize}
\item $\mathcal{M}$ is a \textit{mechanism}, that is any function or algorithm that processes data. In our examples that involve machine learning, $\mathcal{M}$ will be an algorithm that trains a model ($k$-means, SVM).
\item $D$ is a probability distribution on the domain, noted $\domain^*$. A user can draw from $D$ an element and use $\mathcal{M}$ to process it. In our examples, the user can draw a subset $X$ from the training data $\domain^*$, and compute $\M{X}$ that represents the trained model. $X$ is the sensitive data that must be protected.
\item $\rho$ is a function that should be understood as a reward function for the attacker. It measures the closeness between two inputs of $\mathcal{M}$. In our definition, it only outputs two values: $0$ if the two inputs are considered too dissimilar, or $1$ if not. One can define a range of such functions, from strong equality ($\rho(X,X')=1 \iff X=X'$) to weaker matches, like $\rho(X,X')=1 \iff d(X,X)<\varepsilon$ for some distance function $d$ and $\varepsilon>0$.
\item a variable $\delta\in (0,1)$. The quantity $(1-\delta)$ is a bound such that no algorithm can successfully approximate elements drawn from $D$ with probability higher than $(1-\delta)$. The higher $\delta$, the more ``private'' the mechanism is.
\end{itemize}

The mechanism $\mathcal{M}$, the measure function $\rho$ and the distribution $D$ are public and available to the user and the adversary.

\begin{figure}
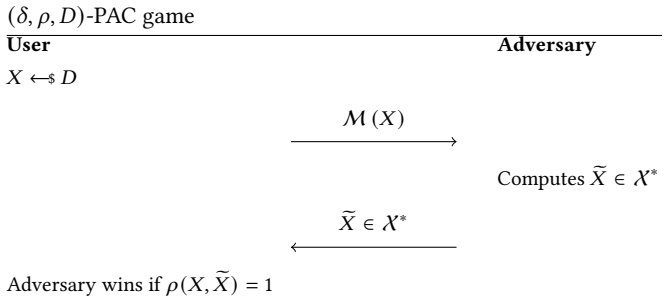

\procedureblock{$(\delta, \rho, D)$-PAC game}{
\textbf{User} \> \> \textbf{Adversary}\\
X\sample D \> \> \\
\> \sendmessageright*[2.2cm]{\M X} \> \\
\> \>\text{Computes }\widetilde{X}\in \domain^* \\
\> \sendmessageleft*[2.2cm]{\widetilde{X}\in \domain^*} \> \\
\text{Adversary wins if } \rho(X,\widetilde{X}) = 1 \> \>}
\caption{PAC game}\label{fig:pac_game}
\end{figure}

A mechanism $\mathcal{M}$ is $(\delta, \rho, D)$-PAC private is there is no adversary able to win the game illustrated in Figure~\ref{fig:pac_game} with probability higher than $(1-\delta)$.
To make a mechanism $\mathcal{M}$ PAC-private, one replaces it with a modified version that has been carefully injected with noise using the presented approach in PAC privacy. We denote by $\B$ the noise distribution added to $\mathcal{M}$.
\cite{xiao2023pac} shows that, given $\delta$ and $\mathcal{M}$, it is always possible to choose $\B$ such that the mutual information between $X$ and $\M{X}+\B$ is small enough to make the noisy mechanism $(\delta, \rho, D)$-PAC private for any function $\rho$. Adding the right amount of noise hence protects $\mathcal{M}(X) + \B$ from leaking any sensitive information. Furthermore, the necessary noise can be small compared to $\M{X}$. For reasonable mechanisms $\mathcal{M}$ whose outputs values are in $\R^d$, it is possible to obtain $\MI{X ; \M X + \B}\leq \beta$ with a Gaussian noise $\B$ that satisfies $\E{\Norm{\B}_2} = O \left(\sqrt{d/\beta}\right)$. This means the noise grows with speed proportional to $\beta^{-1/2}$, where $\beta$ is a bound on $\MI{X ; \M X + \B}$.

\section{Related Works}



Several previous works have motivated research in the realm of verifiable privacy guarantees and verifiable computation based on probabilistic mechanisms (e.g. differential privacy). \cite{narayan2015verifiable} outlines VerDP, a framework that enables private data analysis and queries with differential privacy (DP) while ensuring integrity with ZKPs. This work addresses a core problem with secure computation: proving that computation is done correctly while operating on private data, thus achieving privacy and integrity. VerDP assures that all queries and operations are differentially private, then leverages ZKPs to the integrity of the result, ensuring it was correctly evaluated on the private data without revealing any sensitive information.

Extending this idea to the more complex task of model training, \cite{shamsabadi2024confidential} enables model owners to prove that their models were trained using differentially private stochastic gradient descent (DP-SGD). Their proposed system, Confidential-DPproof, utilizes ZKPs to provide a proof of the privacy guarantee that is ensured by DP, without revealing any information about the training data or model. After training, an auditor can validate this proof to ensure that the model owner did train their model on a private dataset with the promised DP guarantee. This work simplifies the auditing process for correct training of a model, while still ensuring privacy and integrity.

These two works primarily focus on verifying the proper application of differential privacy in real-world systems. \cite{bell2024certifying} takes a different approach by providing proofs that certify the output of a probabilistic mechanism, such as DP. The proposed framework introduces the idea of Certified Probabilistic Mechanisms (CPMs) that allows the verification of a probabilistic mechanism's output without needing to know the mechanism's sensitive parameters. This idea is extended to Certified Differential Privacy (CDP), which allows an auditor to verify that a data curator is releasing information in a way that satisfies DP guarantees.

While all of these works address applications and constructions of verifiably private systems, they primarily focus on DP, a primitive that has been proven to lose utility when applied at scale \cite{blanco2022critical}.
Also, in the case of small datasets, the large noise parameters needed to guarantee a reasonable level of privacy distort the information so much that its exploitation becomes challenging, if not impossible \cite{CriticalDP2025}.
\cite{xiao2023pac} shows that PAC Privacy can lower the amount of necessary noise required to protect sensitive data, when compared to other noise-based privacy-enhancing mechanisms. PAC Privacy utilizes a framework that autonomously determines the minimal noise addition necessary for effective data protection. Rather than generating noise based off a mechanism's inner workings, PAC only relies on black-box access to the mechanism to achieve the same, if not greater, levels of data privacy when compared to DP. While PAC privacy performs well in large-scale systems, providing proofs of integrity that attest to the correct computation and application of noise to protect user data is a challenge, due to the overhead that proper PAC privacy generation requires to achieve high privacy guarantees. Our proposed work is the first to formulate efficient ZKPs that attest maintain integrity and privacy in systems that utilize PAC Privacy.


\section{Methodology}\label{sec:methodology}

\subsection{Threat Model}


Our proposed work is primarily designed to operate in trustless environments, in which all parties must be convinced of computational integrity to a probabilistically high degree. We will assume the following threat model. We first assume a malicious prover, characterized by a server that performs outsourced computation in this work, who may attempt to deviate from the protocol or alter information before the proof is accepted. This malicious assumption ensures that the protocol can withstand potential adversaries who actively seek to corrupt the proof’s integrity. 
We consider the verifier, characterized by a client outsourcing computation, to be semi-honest, following the protocol’s procedures but attempting to deduce as much information as possible from the received data.
This model allows us to analyze both data confidentiality and protocol integrity, ensuring that sensitive information remains protected even against parties that respect the protocol but attempt unauthorized inference.
In a fully malicious setting, one can consider Fiat-Shamir transformations to generate randomness within the circuit, however, in our semi-honest setup, we will consider the verifier's randomness as sufficient.

\subsection{Global Flow}\label{sec:global_flow}
Let $\mathcal{M}$ be a mechanism which can be seen as a function taking a secret input $x$ and producing a public output. As is common in zero-knowledge, an entity called ``Verifier'' will perform a set of computations which are then proven to be correct by the ``Prover''. The goal is to produce a degraded version of $\mathcal{M}(x)$, noted $\mathcal{M}(x) + \B$, where $\B$ is a random vector drawn from a Gaussian distribution and a proof of computation. 

Due to the restricting properties of provable computation in zero-knowledge, we need to work with deterministic functions. Our work introduces two such functions: $f_h$ used for the noise determination and $f_{PAC}$ such that if a data point $x$ is given as input alongside a random seed $s$, the output is an instance of $\mathcal{M}(x) + \B$.
We refer to Algorithms~\ref{algo:general} and \ref{algo:pac} for an illustration of the general flow. 

More precisely, the noise parameters are computed using Algorithm~\ref{algo:general}, adapted from \cite{pacprivatealgos} . They are stored in a Gaussian covariance matrix \Sig. This matrix must remain private, but the proof of its correct computation has to be public and verifiable. That is why the first function $f_h$ takes data points $x_1,\dots,x_m$ as inputs and outputs the hash of \Sig. The hash function $h$ is fixed, public, and supposedly resistant to collisions. That way, it is possible to prove that $h(\Sig)$ really corresponds to a matrix \Sig, and that this matrix is now privately stored by the server.
We propose in Algorithm~\ref{algo:fh} a procedure to generate a noise distribution, and use a zero-knowledge prover to check that the noise is adapted to the mechanism. It is based on the anisotropic noise determination algorithm from \cite{pacprivatealgos}. By theorem 1 of \cite{pacprivatealgos}, the Gaussian noise $\mathcal{B}$ whose covariance matrix is $\Sigma$ is guaranteed to satisfy
$$MI(X;\mathcal{M}(X) + \B) \leq \beta.$$

Once the noise is generated, one can use $f_{PAC}$ to obtain $\mathcal{M}(x)+\B$. The function $f_{PAC}$ takes two private inputs and one public input.
The private inputs are the desired data point $x$, and a random seed $s$. The mechanism $\mathcal{M}$ outputs vectors in $\R^d$ and the seed $s$ is a tuple of $d$ real numbers drawn from a Gaussian normal law. The public input is $h(\Sig):=\widetilde{h}$. That way, it is possible for the server to prove that it really used~\Sig for the computations, by first computing $h(\Sig)$ in zero-knowledge and verifying that $h(\Sig)=\widetilde{h}$. From the seed $s$ and $\Sig$, it is possible to deterministically produce a noise matrix~$\B$ that follows a Gaussian law with covariance matrix~\Sig.

To summarize, we obtain a value for $\mathcal{M}(x) + \B$ with a proof it is correct, without disclosing the parameters of the noise, encoded by~\Sig, that must remain private.

\begin{algorithm}
\caption{Noise generation (follows~\cite[Algorithm 1]{xiao2023pac})}
\label{algo:general}
\textbf{Public Inputs ($\mathcal{P}$ \& $\mathcal{V}$):} function $f_h$  \hfill \

\textbf{Private Inputs ($\mathcal{P}$):} dataset $X$  \hfill \

\vspace{0.5em}
\textbf{Prover} $\mathcal{P}$ \hfill \textbf{Verifier} $\mathcal{V}$
\vspace{0.5em}

\textbf{$\pi \leftarrow$ ZK Circuit:} \hfill \

\begin{algorithmic}[1]
\State \textbf{begin circuit}
\State \textbf{receive} $X = \{x_0, \dots, x_m\}$
\State \textbf{compute and store} covariance matrix $\Sig$
\State \textbf{compute} $f_h(x_0, \dots, x_m) = h(\Sig)$
\State \textbf{end circuit}
\end{algorithmic}

\vspace{0.5em}
\textbf{Send} $h(\Sigma), \pi$ \hfill $\xrightarrow[]{\;\;\;\;\;\;\;\;\;\;\;\;\;\;\;\; h(\Sigma), \pi \;\;\;\;\;\;\;\;\;\;\;\;\;\;\;\;}$ \hfill \textbf{Receive} $h(\Sigma), \pi$

\vspace{0.5em}
\hfill \textbf{Verify proof $\pi$:}

\hfill \textbf{Output:} Accept / Reject
\end{algorithm}

\begin{algorithm}
\caption{PAC main algorithm}
\label{algo:pac}

\textbf{Public Inputs ($\mathcal{P}$ \& $\mathcal{V}$):} function $f_{PAC}$ $h'$ hash of a covariance matrix, datapoint $x$ \hfill \

\textbf{Private Inputs ($\mathcal{P}$):} noise seed $s$ \hfill \

\vspace{0.5em}
\textbf{Prover} $\mathcal{P}$ \hfill \textbf{Verifier} $\mathcal{V}$
\vspace{0.5em}

\textbf{$\pi \leftarrow$ ZK Circuit:} \hfill \

\begin{algorithmic}[1]
\State \textbf{begin circuit}
\State \textbf{receive public} $h'$
\State \textbf{receive secret} $x,s$
\State \textbf{assert} $h' = h(\Sigma)$ \Comment{Proof that the correct $\Sig$ is used}
\State \textbf{output} $f_{PAC}(\underbrace{x,s}_{\mathtt{private}},\underbrace{h(\Sig)}_{\text{public}}) = \mathcal{M}(x) + \B$
\State \textbf{end circuit}
\end{algorithmic}

\vspace{0.5em}
\textbf{Send} $\pi$ \hfill $\xrightarrow[]{\;\;\;\;\;\;\;\;\;\;\;\;\;\;\;\; \pi \;\;\;\;\;\;\;\;\;\;\;\;\;\;\;\;}$ \hfill \textbf{Receive} $\pi$

\vspace{0.5em}
\hfill \textbf{Verify proof $\pi$:}

\hfill \textbf{Output:} Accept / Reject

\end{algorithm}



\begin{algorithm}
\caption{Computation of $f_h$}
\label{algo:fh}
\textbf{Public Inputs ($\mathcal{P}$ \& $\mathcal{V}$):} $y_1,\dots,y_m$ sampled from $\mathcal{M}$, matrix $A$ \hfill \


\vspace{0.5em}
\textbf{Prover} $\mathcal{P}$ \hfill \textbf{Verifier} $\mathcal{V}$
\vspace{0.5em}

\textbf{$\pi \leftarrow$ ZK Circuit:} \hfill \

\begin{algorithmic}[1]
\State \textbf{begin circuit}
\For{$i$ in $1..d$ (dimension of $A$)}
    \State \textbf{compute} $S_i := \{A\cdot y_1, \dots, A\cdot y_m\}$
    \State \textbf{compute} $\sigma_i := $ variance of $S_i$
\EndFor
\State \textbf{compute} $\Sigma[i] = \frac{\sqrt{\sigma_i}\sum_{k=1}^d \sqrt{\sigma_k}}{2\beta}$
\State \textbf{compute} $\Sigma$, diagonal matrix whose $i$-th diagonal coefficient is $\Sigma[i]$
\State \textbf{end circuit}
\end{algorithmic}

\vspace{0.5em}
\textbf{Send} $\Sigma, \pi$ \hfill $\xrightarrow[]{\;\;\;\;\;\;\;\;\;\;\;\;\;\;\;\; \pi, \Sigma \;\;\;\;\;\;\;\;\;\;\;\;\;\;\;\;}$ \hfill \textbf{Receive} $\Sigma, \pi$

\vspace{0.5em}
\hfill \textbf{Verify proof $\pi$:}

\hfill \textbf{Output:} Accept / Reject

\vspace{0.5em}
\textbf{Note:} $\mathcal{P}$ only wins game if $\pi$ is accepted by $\mathcal{V}$. $\pi$ is only accepted if all operations in ZK Circuit are computed soundly and with valid inputs.
\end{algorithm}

\subsection{Choice of mechanisms}
In Section~\ref{sec:global_flow}, we introduced our methodology to prove in zero-knowledge that a given mechanism $\mathcal{M}$ is PAC-private. We now specify which mechanism $\mathcal{M}$ we considered in our study.
\paragraph{$K$-means} The first mechanism considered is the well-known clustering algorithm $K$-means~\cite{lloyd1982least}. This unsupervised algorithm takes as input a dataset of points and outputs $K$ distinct subsets, \emph{i.e.,} clusters, by minimizing the sum of squared distances between each data point and the centroid of its assigned cluster. The algorithm works iteratively by updating these centroids until convergence. This example is also considered in~\cite{xiao2023pac} and to fit the PAC framework, the dataset points are considered private, and the output centroids are public. We refer to Section~\ref{sec:risco_zero_k_means} for details about our implementation of $K$-means and Section~\ref{sec:experimental_results} for the experimental results.

\paragraph{SVMs} The second algorithm considered is the Support Vector Machine (SVM) algorithm. SVM is a supervised learning algorithm used for classification and regression tasks. The algorithm takes as input a dataset and outputs support vectors, \emph{i.e.,} critical data points that lie closest to the decision boundary. More precisely, the algorithm finds a hyperplane that maximally separates data points from different classes, aiming to maximize the margin between the closest points (support vectors) of each class. Similarly as for $K$-means, to fit the PAC framework, we consider the dataset points to be private inputs, and the support vectors to be public. Again, we refer to Section~\ref{sec:risco_zero_k_means} for details about our implementation of SVM and Section~\ref{sec:experimental_results} for the experimental results.
\paragraph{Database statistics} The last mechanism differs from the previous choices that come from machine learning. We focus on queries and statistical operations performed on private dataset points. More precisely, a user wants to acquire the result of statistical operations on parts of a database  that satisfy certain characteristics. The database is privately owned by the server, and the client can only submit queries. These queries act as a filter to select a subset of the database. In this case, a mechanism takes a filter as input and outputs the result of statistical functions (such as the mean, the median etc.) over the filtered points.
Details of our implementation are given in Section~\ref{sec:risc_zero_database} and experimental results are provided in Section~\ref{sec:experimental_results}.



\section{Implementation}
\subsection{Non-Interactive ZK}



Non-interactive ZK (NIZK) generates \textit{publicly-verifiable} proofs by performing this setup process via a trusted third party, or in the case of zk-STARKS, through publicly verifiable randomness. This results in a publicly-available verifier key that can be used to verify the generated proof. This means that one \Prv can generate a proof that can be verified by multiple verifiers, with minimal communication. The downside with NIZKs is their large computational overhead and memory requirements, limiting the scale of applications that can be implemented.
This unfortunately means that in our proposed system, when instantiated in the non-interactive setting with zk-STARKs, the complexity of the PAC-private mechanisms is limited by the ZK scheme itself. 
Our work utilizes RISC-Zero, a state-of-the-art zk-STARK framework, 
to ensure applicability and efficiency in all computational settings.




\begin{figure}
    \centering
    \includegraphics[width=\columnwidth]{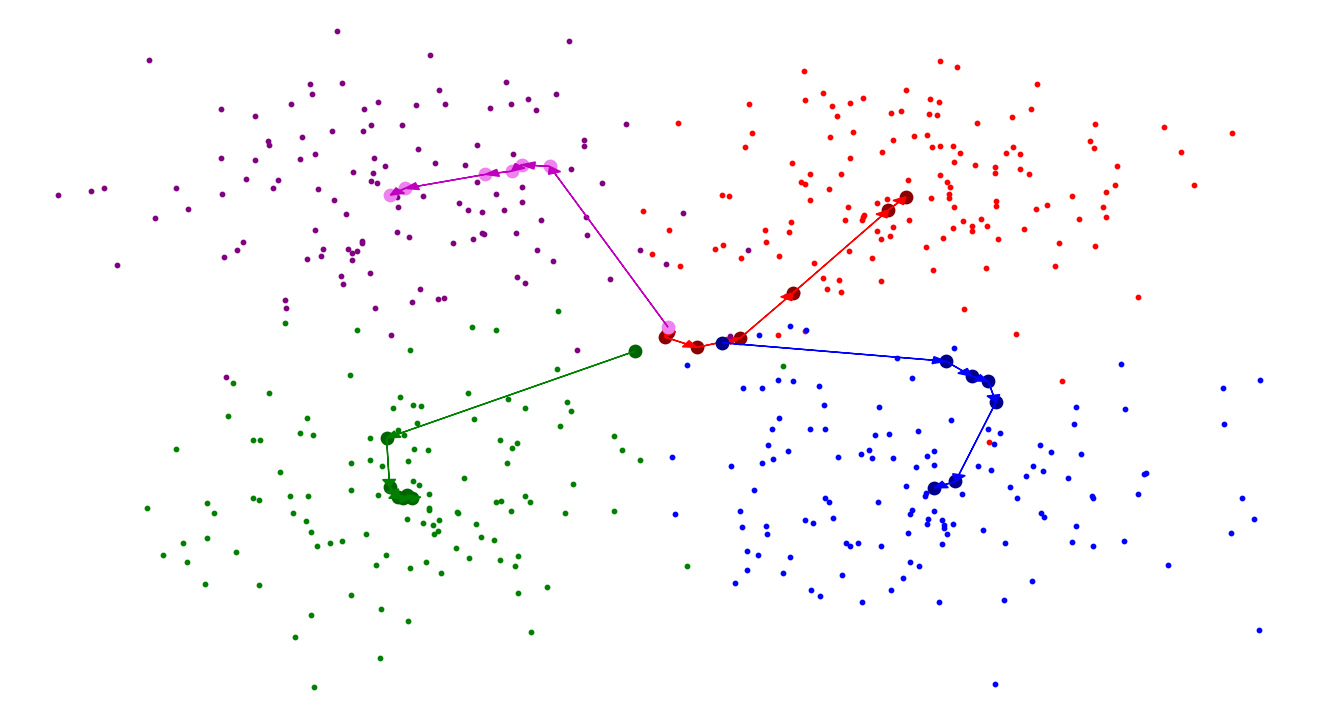}
    \caption{Evolution of the centroids after some iterations of the Risc-Zero implementation of $4$-means ($K=4$)}
    \label{fig:enter-label}
\end{figure}

\subsection{RISC-Zero}
We refer to Section~\ref{sec:global_flow} for the details of the global flow of our algorithm and report here the specifics to the RISC-Zero implementation for each of our chosen mechanisms.

In the context of RISC-Zero, the verifier is called the ``host'' and the prover is called the ``guest''. 
While designing a RISC-Zero program, one must reason as if the guest performed its operations blindly without reading the input, and that the loops are allowed only for concision of the code, but cannot depend on the input $X$.
There are at least two major consequences. One is that it is not possible to compute ``while'' loop, whose number of executions depends on the input $X$. This becomes tricky in the context of machine learning, since most algorithms execute a certain number of rounds until a condition of precision/stability is obtained. The solution we implemented is to execute enough rounds such that the desired condition is probably reached.

The other consequence is that the guest cannot generate randomness, since the program it executes is perfectly deterministic. Again, in the case of algorithms that require a random seed, we face difficulty. The solutions that we implemented to overcome this obstacle is to make the verifier bear the responsibility of the randomness. 

In our case, each mechanism requires two different prover/verifier files: one to generate the noise matrix \B, such as described in Algorithm 1~\cite{pacprivatealgos}, and one to use this prescribed noise distribution to compute a real instance of \M X + \B.

\subsection{ZK Mechanisms}\label{sec:zk_mechanisms}

Let us now revisit the general flow of our zero-knowledge PAC system as described in Section~\ref{sec:global_flow} with our implementation and mechanisms in mind. We will continue to use the verifier/prover terminology to encompass different implementations even though we focus on RISC-Zero in this work.
Implementing a PAC version of an existing mechanism is generally done in two steps:
\begin{itemize}
\item First, one must distinguish the prover's part from the verifier's. The verifier picks a data point $X$ from $\domain^*$, sends it to the prover, whose role is to compute $f(X)$ privately, send back the result to the verifier and publish a proof of correct computation.
\item Second, one must adapt the code so that the execution trace of what the prover does must not depend on the input in receives. This forces the size of $X$ to be fixed, and the potential branchings and loops to always be executed in the same way. These restrictions come from the nature of the proof of execution computed by the prover. It is a proof that the output of the mechanism comes from a known fixed arithmetical circuit. The code executed by the prover is unfolded into such a circuit during compilation, which is a directed acyclic graph (DAG). Once done, this circuit is fixed, its inputs and outputs have a fixed size. The potential loops and tests used can exist only for clarity of the code but not for actual tests. 
\end{itemize}

We now describe our implementations for each of our selected mechanisms.
\subsubsection{$K$-means and SVM}\label{sec:risco_zero_k_means}

Recall that the goal is to classify vectors from $\R^n$ into either $K$ different relevant subgroups for $K$-means or two categories for SVMs. For both algorithms, there is first a training phase of the model, in which a subset $\domain_{\mathrm{train}} \subseteq \domain$ is used to determine automatically the $K$ groups or categories. The $K$-means algorithm requires random numbers, while the SVM algorithm does not.

As mentioned in Devadas and al.~\cite{xiao2023pac}, one chooses a public dataset \domain, and \M{\cdot} trains a $K$-means or SVM model on a subset $\domain_i \subset \domain$ such that $\frac{\Abs{\domain_i}}{\Abs{\domain}} = r$, where $r$ is typically $50\%$.

\paragraph{Noise generation} To generate the adapted noise, in the form of a covariance matrix, the verifier sends random data points to the prover that will use them to compute the covariance matrix.

To compute the noise, the prover executes the mechanism on each of the data-points and then applies the noise generation algorithm from \cite{pacprivatealgos}.

Note that in the case of $K$-means, all the randomness comes from the input.
We now list the main changes made to the standard algorithms to fit our zero-knowledge context.

\textbf{For $K$-means:}
\begin{itemize}
\item $K$ random points are chosen to be the initial centroids. In the zero-knowledge version, the first $K$ points of the input $X$ are chosen to be the centroids ;
\item At first all points are randomly assigned a centroid. In the zero-knowledge version, the points $X[0], X[K], X[2K]\dots$ are assigned to centroid $0$, $X[1], X[K+1], X[2K+1]\dots$ are assigned to centroid $1$, and so on. A list variable \texttt{groups} is used, such that \texttt{groups[i]} contains all vectors assigned to centroid \texttt{i}. Since all the \texttt{groups[i]} must have a fixed size no matter the input, the following trick is used: each \texttt{groups[i]} is a list type $(\texttt{point}, \texttt{bool})$. The list has a fixed size equal to the number of samples, and is used to store points and a boolean indicating if the associated point is to be taken into account.
For instance, if \texttt{groups[i] = [(a, True), (b, False), (c, True)]}, then at this moment in the algorithm, only \texttt{a} and \texttt{c} are assigned to centroid \texttt{i}.

That way, even when the number of points associated to a centroid $i$ varies during the execution of the algorithm, the size of \texttt{groups[i]} stays constant, which is unavoidable in this context of arithmetical circuits.
\item The standard algorithm iterates until a fixed point is attained. In the zero-knowledge setup, the number of iteration is fixed.
\item As the standard version of $\text{$K$-means}(\domain_i)$ returns the list of centroids in any order, a canonicalization method, as discussed in \cite{xiao2023pac}, should be implemented. Here, the centroids are sorted so that the $i$-th represents class $i$. This is done by inferring the class of a centroid. We tried sorting them using the lexicographic order, but it does not work in high dimension, as it is most of the time equivalent to sorting based on the first coordinate.
\end{itemize}
Figure~\ref{fig:enter-label} illustrates the evolution of the centroids after seven iterations of the Risc-Zero implementation for $K=4$.\\

\textbf{For SVMs:}
Since the algorithm iterates over all the points of $\domain_{\mathrm{train}}$ for each epoch, and that the model solely consists of a vector $w$ and a real number $b$, such that the hyperplane is the points satisfying the equation $$ w \cdot z + b = 0,$$
there is less need for adaptation. The only relevant detail concerns the canonicalization of the result. As the noise \B is computed using the average of many SVM models $(w_i, b_i)$, it is crucial that those models are of the same scale: if a point $z$ satisfies $w \cdot z + b = 0$, then it also satisfies $(\lambda w) \cdot z + \lambda b = 0$ for all real number $\lambda$. 

As a result, all models $(w,b)$ are normalized such that forall $i,j$, $\Norm{w_i}=1$, and $w_i \cdot w_j \geq 0$.

To summarize, \domain is either a $K$-labeled-dataset (respectively a $2$-labeled-dataset), with an order on the labels, and $\text{$K$-means}(\domain_i)$ (resp. $\text{SVM}(\domain_i)$) returns $k$ centroids corresponding to those $k$ labels in the given order (resp. a separating hyperplane):

$$
\mathcal{M}_{\text{$K$-means}}:
\begin{matrix}
\left\{ S \subset \domain, \frac{\Abs{S} }{\Abs{\domain} } = r \right\} & \longrightarrow & (\R^n)^k \\
\domain_i  & \longmapsto & \text{canonicalized $k$-means}(\domain_i) \\
\end{matrix}
$$

$$
\mathcal{M}_{\text{SVM}}:
\begin{matrix}
\left\{ S \subset \domain, \frac{\Abs{S} }{\Abs{\domain} } = r \right\} & \longrightarrow & \R^n \times \R \\
\domain_i  & \longmapsto & \text{canonicalized svm}(\domain_i) \\
\end{matrix}
$$

\paragraph{PAC model} Now that we have privately generated \B, we can produce an actual PAC-private $K$-means (resp. SVM) model.
The verifier sends $X$ for which it wants $K$-means (resp. SVM) to be computed on, alongside a secret random normal Gaussian vector $s$, and the hash of the covariance matrix $h(\Sig)$. 
The prover proves that their the hash of their covariance matrix is $h(\Sig)$, and then computes a random vector $v$ from distribution \B using $s$, applies $K$-means (resp. SVM) to $X$, and sends back the sum of the two.




\subsubsection{Database statistics}\label{sec:risc_zero_database}

In this case, $\mathcal{D}$ is a database, and \domain is the powerset of $\mathcal{D}$. A database ia a set of datapoints which are a tuple of \textbf{Attribute}. For instance: 
\begin{center}
\begin{tabular}{|l|c|r|}
\hline
\textbf{Name} & \textbf{Age} & \textbf{Wealth}\\
\hline
Marty & 17 & 10000\\
\hline
Emmett & 65 & 140000 \\
\hline
Biff & 19 & 6000 \\
\hline
Lorraine & 47 & 50000 \\
\hline
\end{tabular}
\end{center}

We define a small set of queries with the following syntax:
\begin{center}
 $q=$``\underline{function} of \textbf{Attribute} \textit{with} $\varphi$''
\end{center}
with \underline{function} $\in \{$median, average, \dots\ $\}$. This finite set of functions can contain any function of type $\textbf{Attribute}\rightarrow \R^m$. Here $\varphi$ is a ``filter'' that defines the set of datapoints the user wants the \underline{function} to be applied on. A filter is constructed recursively the following way:
\begin{equation}
\label{eq:recur}
 \varphi = \varnothing \;|\; \textbf{Attribute} > x \;|\; \textbf{Attribute} = x \;|\; \NOT\, \varphi \;|\; \varphi\, \AND\, \varphi
\end{equation}

Here is an example of query:
\begin{center}
 $q_1=$``\underline{average} of \textbf{Ages} \textit{with} Wealth$>$11000''
\end{center}


In this context, a mechanism is defined by a tuple $T= ($\underline{function 1} of \textbf{Attribute 1},\dots, \underline{function $n$} of \textbf{Attribute $n$}$)$, and $\mathcal{M}_T$ takes a filter $\varphi$ as an input, and returns the vector
$$\begin{bmatrix}
\text{\underline{function 1} of \textbf{Attribute 1} \textit{with }} \varphi \\
\vdots \\
\text{\underline{function n} of \textbf{Attribute n} \textit{with }} \varphi
\end{bmatrix}$$

Formally:
$$\mathcal{M}_T : \begin{array}{lll}
 & \text{Set of filters} &\longrightarrow \R^n \\
     &\varphi &\longmapsto \begin{bmatrix}
\text{\underline{function 1} of \textbf{Attribute 1} \textit{with }} \varphi \\
\vdots \\
\text{\underline{function n} of \textbf{Attribute n} \textit{with }} \varphi
\end{bmatrix}
\end{array}$$

For instance: if $T= ($\underline{average} of \textbf{Age}, \underline{median} of \textbf{Wealth}$)$, and $\varphi = \textbf{Age}>25$, then
$$M_T (\varphi) = \begin{bmatrix}
\text{\underline{average} of \textbf{Age} \textit{with }} \textbf{Age}>25 \\
\text{\underline{median} of \textbf{Wealth} \textit{with }} \textbf{Age}>25
\end{bmatrix} = \begin{bmatrix}
37 \\
200000
\end{bmatrix}$$

\paragraph{Noise generation} As per the PAC noise generating algorithm, this first stage needs to generate data in order to get a noise matrix. This data is composed a collection of filters $\varphi^{(1)},\dots, \varphi^{(m)}$.

Recall that zero-knowledge programming adds the constraint that all the $\varphi^{(i)}$ must occupy a fixed  size in memory. To remedy this issue, we use a special \texttt{Formula} type, which is simply a vector of length twice the number of attributes. The semantics of such a \texttt{Formula} \texttt{f} is 
\begin{align*}
\varphi = &\texttt{f}[0] \leq \textbf{Attribute}[0] \, \AND \, \textbf{Attribute}[0] < \texttt{f}[1] \\
\AND \, &\texttt{f}[2] \leq \textbf{Attribute}[1] \, \AND \, \textbf{Attribute}[1] < \texttt{f}[3] \\
\vdots \\
\AND \, &\texttt{f}[2k] \leq \textbf{Attribute}[k] \, \AND \, \textbf{Attribute}[k] < \texttt{f}[2k+1]
\end{align*}

For instance, in our example, $f=\texttt{[26,100,0,100000]}$ designates the datapoints satisfying $26\leq\mathbf{Age}<100$ and $0\leq \mathbf{Wealth}<100000$.

The chosen method to generate a random filter is to a random \texttt{Formula} is to draw uniformly each coordinates on its corresponding span. For instance, in our example the bounds for \textbf{Age} are drawn uniformly from $[\![24,51]\!]$.

Another unexplored way could be to compute small decision trees in order to split the database into relevant subparts, and creation of filters corresponding to each of their leaves.

As in $K$-means, the verifier sends a list of filters \texttt{f1,\dots,fn}. The prover iterates on the database to find all points satisfying the conditions described by each \texttt{fi}. The noise matrix \B is then generated by applying $\mathcal{M}_T$.


\section{Experimental Evaluation}\label{sec:experimental_results}

\subsection{Experimental Setup}
All the mechanisms above are implemented in the Risc0 framework. They are all executed on small datasets. By variying a parameter on the dataset (e.g. dimension of the points) and on the mechanism (e.g. number of samples $M$, number of clusters for $K$-means), we observe the evolution of the execution time of the noise generation algorithm. The latter is measured in terms of number of simulated Risc operations, and is thus independant from the actual used machine. In each case, we expect an affine relation in the number of samples, since a major part of the noise generating algorithm involves looping over each sample.
\subsection{Experimental Results}

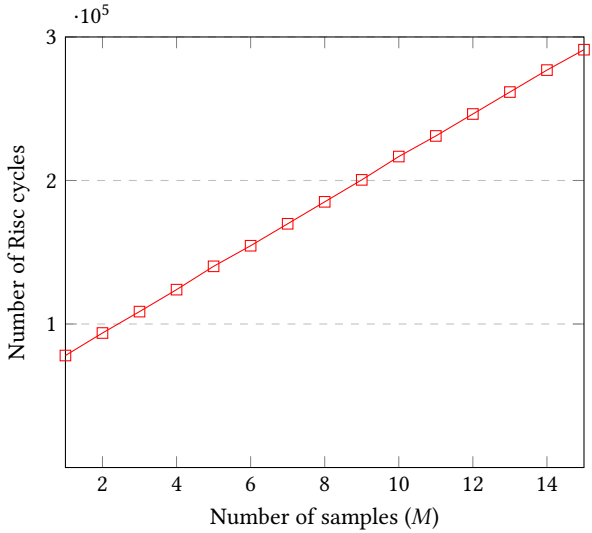
\begin{figure}
    \centering
\begin{tikzpicture}
\begin{axis}[
    xlabel={Number of samples ($M$)},
    ylabel={Number of Risc cycles},
    xmin=1, xmax=15,
    ymin=0, ymax=300000,
    xtick={0,2,4,6,8,10,12,14},
    ytick={100000,200000,300000},
    legend pos=north west,
    ymajorgrids=true,
    grid style=dashed,
]

\addplot[
    color=red,
    mark=square,
    ]
    coordinates {
    (1,78046)(2,93727)(3,108642)(4,123940)(5,140195)(6,154536)(7,169834)(8,185132)(9,200430)(10,216685)(11,231026)(12,246324)(13,261622)(14,276920)(15,291130)
    };
    \label{p13}

\end{axis}
\end{tikzpicture}
\caption{Execution $K$-means with different number of samples. The database consists of $1000$ points, and $500$ random points are used for each training.}
\label{fig:timing}
\end{figure}

\begin{figure}
    \centering
\begin{tikzpicture}
\begin{axis}[
    xlabel={$K$},
    ylabel={Number of Risc cycles},
    xmin=2, xmax=10,
    ymin=0, ymax=1400000000,
    xtick={0,1,2,3,4,5,6,7,8,9,10},
    ytick={0,300000000,600000000,900000000,1200000000},
    legend pos=north west,
    ymajorgrids=true,
    grid style=dashed,
]

\addplot[
    color=blue,
    mark=o,
    ]
    coordinates {
    (2,135074907)(3,193708607)(4,251629736)(5,309215292)(6,366651939)(7,423592173)(8,479114612)(9,536860996)(10,594068350)
    };
    \label{p21}
    
\addplot[
    color=cyan,
    mark=*,
    ]
    coordinates {
    (2,269867679)(3,387263881)(4,504078762)(5,619759345)(6,736082496)(7,850337172)(8,967693827)(9,1079906999)(10,1195361605)
    };
    \label{p22}

\addplot[
    color=red,
    mark=square,
    ]
    coordinates {
    (2,450864248)(3,640087472)(4,829513513)(5,1019219914)(6,1208293866)(7,1397902515)
    };
    \label{p23}
    
\addplot[
    color=magenta,
    mark=square*,
    ]
    coordinates {
    (2,225845154)(3,320559015)(4,415534080)(5,510316711)(6,604930705)(7,699646700)(8,794072321)(9,887790937)(10,982003271)
    };
    \label{p24}


\node [draw,fill=white] at (rel axis cs: 0.75,0.15) {\shortstack[l]{
\ref{p21} 5 iterations, $d=2$ \\
\ref{p22} 10 iterations, $d=2$ \\
\ref{p23} 10 iterations, $d=4$ \\
\ref{p24} 5 iterations, $d=4$
}};
\end{axis}
\end{tikzpicture}
\caption{Execution $K$-means with different values of $K$, and dimensions of the points, and number of iterations in the training phase. The database consists of $1000$ points, and $500$ random points are used for each training.}
\label{fig:plot2}
\end{figure}
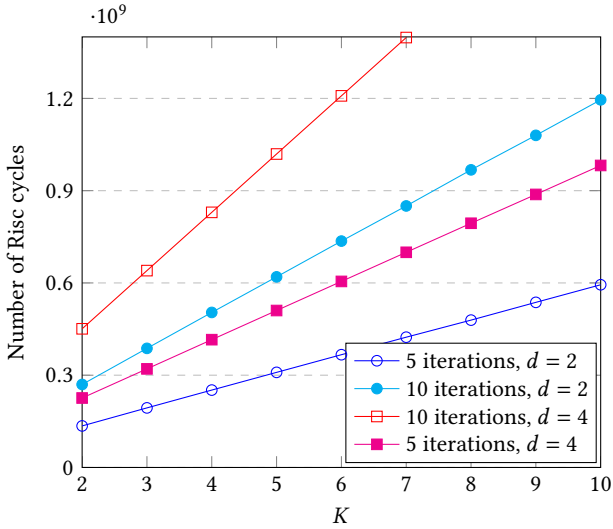

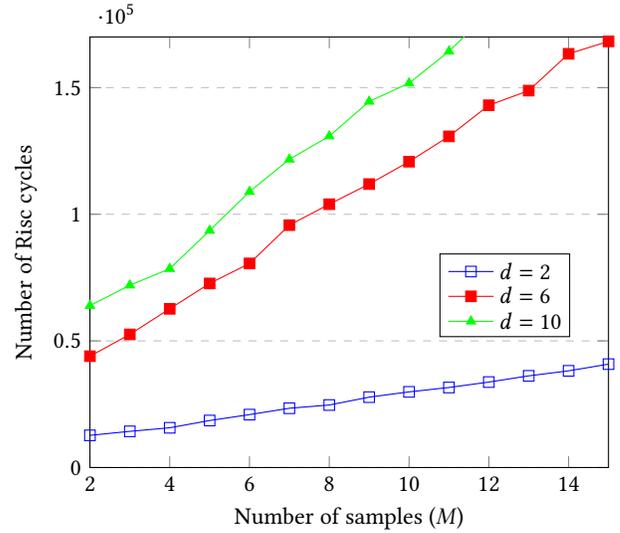
\begin{figure}
    \centering
\begin{tikzpicture}
\begin{axis}[
    xlabel={Number of samples ($M$)},
    ylabel={Number of Risc cycles},
    xmin=2, xmax=15,
    ymin=0, ymax=170000,
    xtick={0,2,4,6,8,10,12,14},
    ytick={0,50000,100000,150000},
    legend pos=north west,
    ymajorgrids=true,
    grid style=dashed,
]

\addplot[
    color=blue,
    mark=square,
    ]
    coordinates {
    (2,12702)(3,14283)(4,15692)(5,18554)(6,20892)(7,23401)(8,24685)(9,27791)(10,29872 )(11,31602)(12,33743)(13,36196)(14,38180)(15,40826)
    };
    \label{p31}

\addplot[
    color=red,
    mark=square*,
    ]
    coordinates {
    (2,43948)(3,52564)(4,62669)(5,72674)(6,80572)(7,95701)(8,103973)(9,111918)(10,120767)(11,130744)(12,143018)(13,148810)(14,163350)(15,168260)
    };
    \label{p32}

\addplot[
    color=green,
    mark=triangle*,
    ]
    coordinates {
    (2,63921)(3,71963)(4,78483)(5,93562)(6,108906)(7,121682)(8,130838)(9,144513)(10,151734)(11,164324)(12,179553)
    };
    \label{p33}

\node [draw,fill=white] at (rel axis cs: 0.8,0.4) {\shortstack[l]{
\ref{p31} $d=2$ \\ 
\ref{p32} $d=6$ \\
\ref{p33} $d=10$
}};
\end{axis}
\end{tikzpicture}
\caption{Number of Risc0 cycles needed to compute the noise generating algorithm for SVM. Each sample consists of $50$ datapoints are of dimensions $2,6$ and $10$. The SVM algorithm executes $1000$ epochs.}
\label{fig:svm_timing}
\end{figure}

\paragraph{K-means:} Figure~\ref{fig:timing} shows the number of cycles of simulated Risc-Zero processor needed to execute $K$-means, with $K=2$. Figure~\ref{fig:plot2} shows the number of cycles needed for various values of $K$. They display a perfect affine growth of the number of Risc operations in the number of samples ($M$) and in $K$. This was expected, since the algorithm loops both over those two quantities, and each round executes exactly the same number of cycles due to the deterministic nature of the computation. In other words, if the dimension $d$ of the points is fixed, the number of cycles is an affine function of $K$. If $K$ is fixed, the number of cycles is an affine function of the dimension $d$. Once the noise is generated, the number of cycles needed to execute the PAC version of $K$-means is substantially the same as the value corresponding to one sample in the chart.
\paragraph{SVM:} For this mechanism, we do not observe a perfect affine relation (see Figure \ref{fig:svm_timing}). The reason is that the SVM algorithm stores more data than $K$-means, and hence needs the Risc0 processor to use the cache. Slight performance fluctuations for the SVM algorithms could come from memory-intensive operations than with K-means in the zkVM circuit, therefore leading to some variation in the cycle counts. Caching operations last a non-predictable number of cycles, which varies. We still observe a general affine behavior in the number of samples ($M$).

\begin{figure}
\centering
\begin{tikzpicture}[]
\begin{axis}[
boxplot/draw direction=y,
xmin=25, xmax=1075,
ymin=0, ymax=440000,
xtick={0,100,200,300,400,500,600,700,800,900,1000},
ytick={0,100000,200000,300000,400000},
     xlabel={Number of points in the database},
     ylabel={Number of Risc cycles},
]


\addplot+ [solid, red, boxplot prepared={box extend=2*\gap,draw position=1000+\gap, lower whisker=335183, lower quartile=354771, median=367717, upper quartile=379780, upper whisker=398275}, ] coordinates {};

\addplot+ [solid, red,  boxplot prepared={box extend=2*\gap,draw position=900+\gap, lower whisker=303057, lower quartile=325392, median=335475, upper quartile=361787, upper whisker=401949}, ] coordinates {};

\addplot+ [solid, red, boxplot prepared={box extend=2*\gap,draw position=800+\gap, lower whisker=271933, lower quartile=295017, median=300046, upper quartile=309714, upper whisker=330981}, ] coordinates {};

\addplot+ [solid, red, boxplot prepared={box extend=2*\gap,draw position=700+\gap, lower whisker=239595, lower quartile=257778, median=268139, upper quartile=273580, upper whisker=294302}, ] coordinates {};

\addplot+ [solid, red, boxplot prepared={box extend=2*\gap,draw position=600+\gap, lower whisker=215779, lower quartile=225072, median=232168, upper quartile=240679, upper whisker=262454}, ] coordinates {};

\addplot+ [solid, red, boxplot prepared={box extend=2*\gap,draw position=500+\gap, lower whisker=167361, lower quartile=178185, median=183029, upper quartile=188085, upper whisker=212783}, ] coordinates {};

\addplot+ [solid, red, boxplot prepared={box extend=2*\gap,draw position=400+\gap, lower whisker=138321, lower quartile=144682, median=150062, upper quartile=153639, upper whisker=173958}, ] coordinates {};

\addplot+ [solid, red, boxplot prepared={box extend=2*\gap,draw position=300+\gap, lower whisker=102466, lower quartile=112935, median=117189, upper quartile=123526, upper whisker=143154}, ] coordinates {};

\addplot+ [solid, red, boxplot prepared={box extend=2*\gap,draw position=200+\gap, lower whisker=69427, lower quartile=75380, median=77742, upper quartile=81492, upper whisker=92512}, ] coordinates {};

\addplot+ [solid, red, boxplot prepared={box extend=2*\gap,draw position=100+\gap, lower whisker=36638, lower quartile=40938, median=42687, upper quartile=44338, upper whisker=48832}, ] coordinates {};


\addplot+ [solid, blue, boxplot prepared={box extend=2*\gap,draw position=1000-\gap, lower whisker=373921, lower quartile=378146, median=382793, upper quartile=388974, upper whisker=394974}, ] coordinates {};

\addplot+ [solid, blue,  boxplot prepared={box extend=2*\gap,draw position=900-\gap, lower whisker=334702, lower quartile=341267, median=347006, upper quartile=350940, upper whisker=356287}, ] coordinates {};

\addplot+ [solid, blue, boxplot prepared={box extend=2*\gap,draw position=800-\gap, lower whisker=301958, lower quartile=304876, median=310547, upper quartile=312803, upper whisker=321541}, ] coordinates {};

\addplot+ [solid, blue, boxplot prepared={box extend=2*\gap,draw position=700-\gap, lower whisker=268658, lower quartile=272259, median=275588, upper quartile=278509, upper whisker=279622}, ] coordinates {};

\addplot+ [solid, blue, boxplot prepared={box extend=2*\gap,draw position=600-\gap, lower whisker=231952, lower quartile=236606, median=237024, upper quartile=239295, upper whisker=244447}, ] coordinates {};

\addplot+ [solid, blue, boxplot prepared={box extend=2*\gap,draw position=500-\gap, lower whisker=184891, lower quartile=190675, median=193216, upper quartile=195031, upper whisker=196261}, ] coordinates {};

\addplot+ [solid, blue, boxplot prepared={box extend=2*\gap,draw position=400-\gap, lower whisker=152364, lower quartile=154937, median=156286, upper quartile=159531, upper whisker=165623}, ] coordinates {};

\addplot+ [solid, blue, boxplot prepared={box extend=2*\gap,draw position=300-\gap, lower whisker=117499, lower quartile=120722, median=122039, upper quartile=124755, upper whisker=129439}, ] coordinates {};

\addplot+ [solid, blue, boxplot prepared={box extend=2*\gap,draw position=200-\gap, lower whisker=78035, lower quartile=79896, median=80925, upper quartile=82186, upper whisker=85512}, ] coordinates {};

\addplot+ [solid, blue, boxplot prepared={box extend=2*\gap,draw position=100-\gap, lower whisker=41976, lower quartile=43245, median=43566, upper quartile=44560, upper whisker=47315}, ] coordinates {};

\end{axis}

\node [draw,fill=white] at (rel axis cs: 0.8,0.1) {\shortstack[l]{
\color{blue}$\blacksquare$ \color{black} $d=4$  \\
\color{red}$\blacksquare$ \color{black} $d=2$ 
}};

\end{tikzpicture}
\caption{Number of Risc0 cycles to compute one PAC-private query. The dataset consists of points of dimension $d$. The quartiles, median, maximum and minimum are computed from $50$ random queries. }\label{fig:database}
\end{figure}
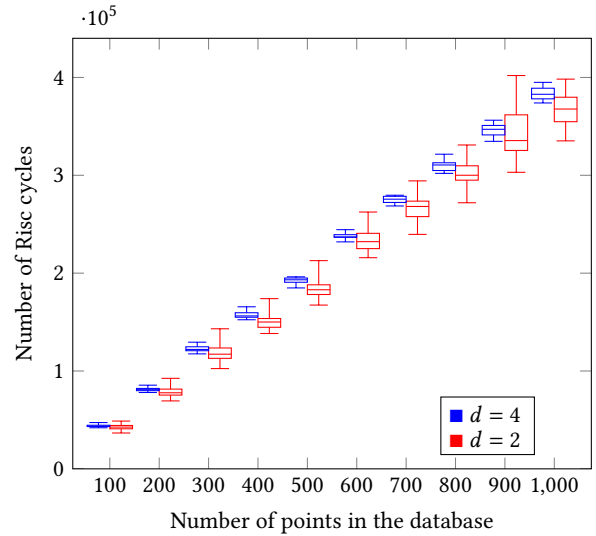

\paragraph{Database statistics:} Finally, Figure \ref{fig:database} shows the same affine behavior. The results are obtained by timing the query mechanism over random queries, for different database sizes and point dimensions.

\tikzmath{\gap = 23;}

\section{Conclusion}

This paper introduced the first framework that combines PAC Privacy with zk-STARK-based zero-knowledge proofs to provide verifiable privacy in outsourced computations. By instantiating the framework in RISC-Zero, we showed that PAC privacy can be enforced with proofs of correctness without revealing raw data or noise parameters. We evaluated the approach on K-means, SVM, and statistical queries, finding that proof overhead scales predictably with dataset size and remains feasible for small to medium applications.

 \begin{acks}
We would like to thank Prof. Srini Devadas and Prof. Hanshen Xiao for their fruitful discussions about this work and their valuable feedback.
\end{acks}

\bibliographystyle{ACM-Reference-Format}
\bibliography{refs}

\appendix
\section{Mathematical background}

\subsection{Multimodal Gaussian noise generation}
The PAC algorithm must generate a multimodal Gaussian noise out of a covariance matrix \Sig and a random vector $\mathbf{Z}$ whose coordinates are reduced centered laws. For that, a solution that is implementable in zero-knowledge is to compute the Choleski decomposition of \Sig.

\begin{lemma}[Choleski decomposition]
If $\Sigma$ is symmetric definite-positive, then there exists a unique real matrix $A$ that is lower triangular with positive diagonal entries, such that
$$\Sigma = AA^T$$
\end{lemma}



\begin{lemma}[Noise generation]

From a random vector sampled from $\mathcal{N}(0_n, I_n)$, we can obtain a vector sampled from $\mathcal{N}(\mu, \Sigma)$, using the Choleski decomposition of $\Sigma$:

$$\B \sim \mathcal{N}(\mu, \Sigma) \quad \Longleftrightarrow \quad \left\{
\begin{matrix}
 \B & = & A \mathbf{Z} + \mu \\
 \forall j, \mathbf{Z}_j & \sim & \mathcal{N}(0, 1)\\
 AA^T & = &\Sigma
\end{matrix}
\right.$$

In particular:
$$\B \sim \mathcal{N}(0, \Sigma) \quad \Longleftrightarrow \quad  \B = A \mathbf{Z}$$
where $\mathbf{Z} \sim \mathcal{N}(0, I_n)$ and $A$ is the Choleski decomposition of $\Sigma$.
\end{lemma}

\end{document}